\documentclass[lettersize,journal]{IEEEtran}
\usepackage{amsmath,amsfonts}
\usepackage{algorithmic}
\usepackage{algorithm}
\usepackage{array}
\usepackage[caption=false,font=normalsize,labelfont=sf,textfont=sf]{subfig}
\usepackage{textcomp}
\usepackage{stfloats}
\usepackage{url}
\usepackage{verbatim}
\usepackage{graphicx}
\usepackage{cite}
\usepackage{tikz}
\usetikzlibrary{arrows}
\usepackage{amssymb}

\begin{document}

\title{Hybrid Table-Assisted and RL-Based Dynamic Routing for NGSO Satellite Networks}

\author{Flor Ortiz, and Eva Lagunas
\thanks{The authors are with the Interdisciplinary Centre for Security, Reliability, and Trust (SnT), University of Luxembourg, 1855 Luxembourg City, Luxembourg.
(email:{flor.ortiz, eva.lagunas}@uni.lu).}
\thanks{This research was funded in whole, or in part, by the Luxembourg National Research Fund (FNR), grant reference [DEFENCE22/
17801293/GSDT].}}

\markboth{Journal of \LaTeX\ Class Files,~Vol.~14, No.~8, August~2021}%
{Shell \MakeLowercase{\textit{et al.}}: A Sample Article Using IEEEtran.cls for IEEE Journals}


\maketitle

\begin{abstract}
This letter investigates dynamic routing in Next-Generation Satellite Orbit (NGSO) constellations and proposes a hybrid strategy that combines precomputed routing tables with a Deep Q-Learning (DQL) fallback mechanism. While fully RL-based schemes offer adaptability to topology dynamics, they often suffer from high complexity, long convergence times, and unstable performance under heavy traffic. In contrast, the proposed framework exploits deterministic table lookups under nominal conditions and selectively activates the DQL agent only when links become unavailable or congested. Simulation results in large-scale NGSO networks show that the hybrid approach consistently achieves higher packet delivery ratio, lower end-to-end delay, shorter average hop count, and improved throughput compared to a pure RL baseline. These findings highlight the effectiveness of hybrid routing as a scalable and resilient solution for delay-sensitive satellite broadband services.
\end{abstract}


\section{Introduction}
\IEEEPARstart{N}{ext}-Generation Satellite Orbit (NGSO) constellations are a key enabler of global broadband, offering low-latency connectivity through dense Low Earth Orbit (LEO) deployments \cite{9460990}. Unlike geostationary systems, NGSO networks exhibit rapidly time-varying topologies in which inter-satellite and feeder links change with orbital motion and ground visibility, so static routes become quickly outdated and degrade performance \cite{11062652}.

Reinforcement learning (RL) has been explored to endow satellites with adaptive next-hop decisions based on local observations (e.g., link availability, queue states, delay estimates) \cite{10436098}. Fully RL-driven routing, however, may suffer from long convergence, route instability during training, and non-trivial onboard complexity; under nominal conditions, lightweight table-based methods can be more efficient \cite{HUANG2023284}. In current practice, NGSO routing often relies on pre-configured tables derived from packet/header inspection; Multiprotocol Label Switching (MPLS) is a representative mechanism where labels are looked up in static tables to select the next hop \cite{10.1108/978-1-83797-052-020241016}. Table-based forwarding is simple and stable but lacks flexibility under congestion, intermittent connectivity, or sudden traffic shifts. These complementary strengths motivate a hybrid design that couples deterministic lookups with selective learning-based adaptation.

\subsection{Related Work}
Conventional NGSO strategies implement shortest-path routing or MPLS-like forwarding from offline topology knowledge, forwarding packets according to static routes or preconfigured label tables \cite{10.1108/978-1-83797-052-020241016}. While low-overhead and robust in steady conditions, they become suboptimal as visibility and load evolve.

Learning-based methods address this limitation. A distributed Q-learning framework formulated routing as a POMDP, reducing signaling while achieving latencies comparable to shortest-path baselines; yet tabular Q-learning struggles with large state/action spaces and heavy congestion \cite{10624807}. MA-DRL extensions improved adaptability by combining offline exploration with online exploitation, at the cost of pre-trained global models and higher onboard computation \cite{10624767}. Continual learning mitigates catastrophic forgetting as traffic patterns evolve, but retains substantial training overhead and may converge unstably under rapidly fluctuating loads \cite{10969776}. Beyond routing, MA-DRL has been used for mobility management and handovers in mega-constellations, improving service continuity but not directly optimizing end-to-end routing \cite{10551688}. The QRLSN scheme integrates RL with multi-objective optimization, outperforming virtual-topology shortest paths in QoS, while incurring significant training time and reduced efficiency at light loads \cite{HUANG2023284}. Broader surveys synthesize these trade-offs—dynamic topology, latency dominance, and limited onboard resources—highlighting the value of hybrid solutions that balance deterministic efficiency with adaptive decision-making \cite{11000293}.
\subsection{Contributions}

Motivated by the limitations of fully RL-based routing approaches and the shortcomings of purely static methods, this letter introduces a nobel hybrid strategy that combines deterministic routing tables with a Deep Q-Learning (DQL) fallback mechanism. The proposed solution aims to provide both efficiency under nominal conditions and adaptability under disruptions. The main contributions are summarized as follows:

\begin{itemize}
    \item We design a dual-mode routing framework for NGSO constellations that exploits precomputed routing tables for low-complexity forwarding and activates a DQL agent only when static routes are unavailable due to congestion or link failures.
    \item We model the routing problem as a dynamic optimization task over a time-varying graph, explicitly capturing end-to-end delay, queuing effects, and link availability constraints, which provides a principled foundation for the hybrid approach.
    \item Unlike fully RL-based strategies that continuously require training and exploration, our method limits the use of DQL to exceptional events, thereby reducing convergence time, exploration overhead, and routing instability.
    \item Through simulations of large-scale NGSO constellations, we benchmark the hybrid approach against a fully RL-based baseline. Results show that the proposed method achieves consistently lower end-to-end delay, higher packet delivery ratio, and reduced routing overhead, while retaining resilience under dynamic traffic and link conditions.
\end{itemize}

\section{System Model and Problem Formulation}
\begin{figure*}[t]
    \centering
    \subfloat[UT/GW geographical deployment. Stars denote user terminals (UTs) and triangles denote gateways (GWs).]{%
        \includegraphics[width=.4\linewidth]{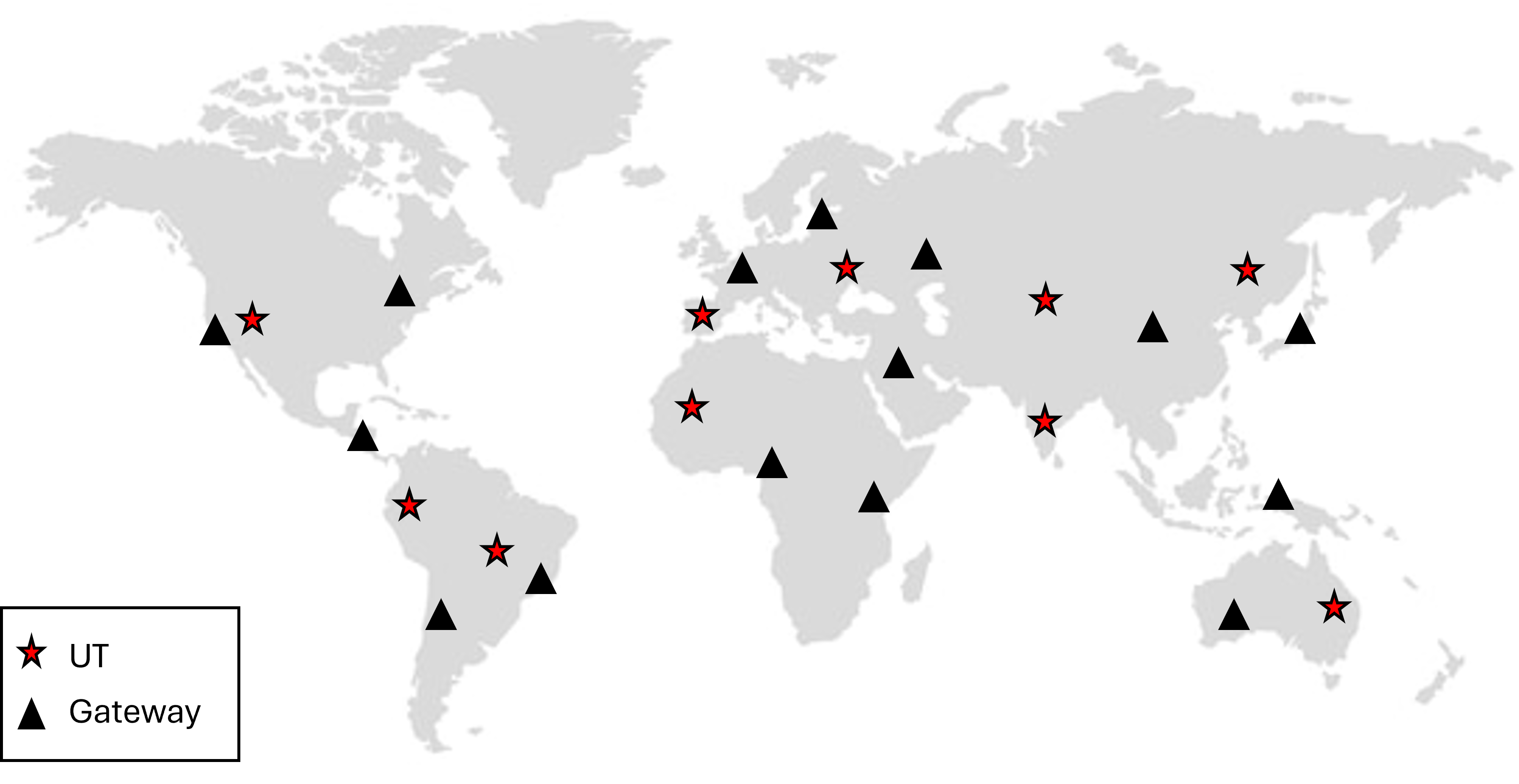}%
        \label{fig:system_a}
    }\hfill
    \subfloat[LEO Walker constellation with intra–/inter–plane ISLs. Green arrows: intra–plane (east/west). Blue arrows: inter–plane (forward/backward).]{%
        \includegraphics[width=.4\linewidth]{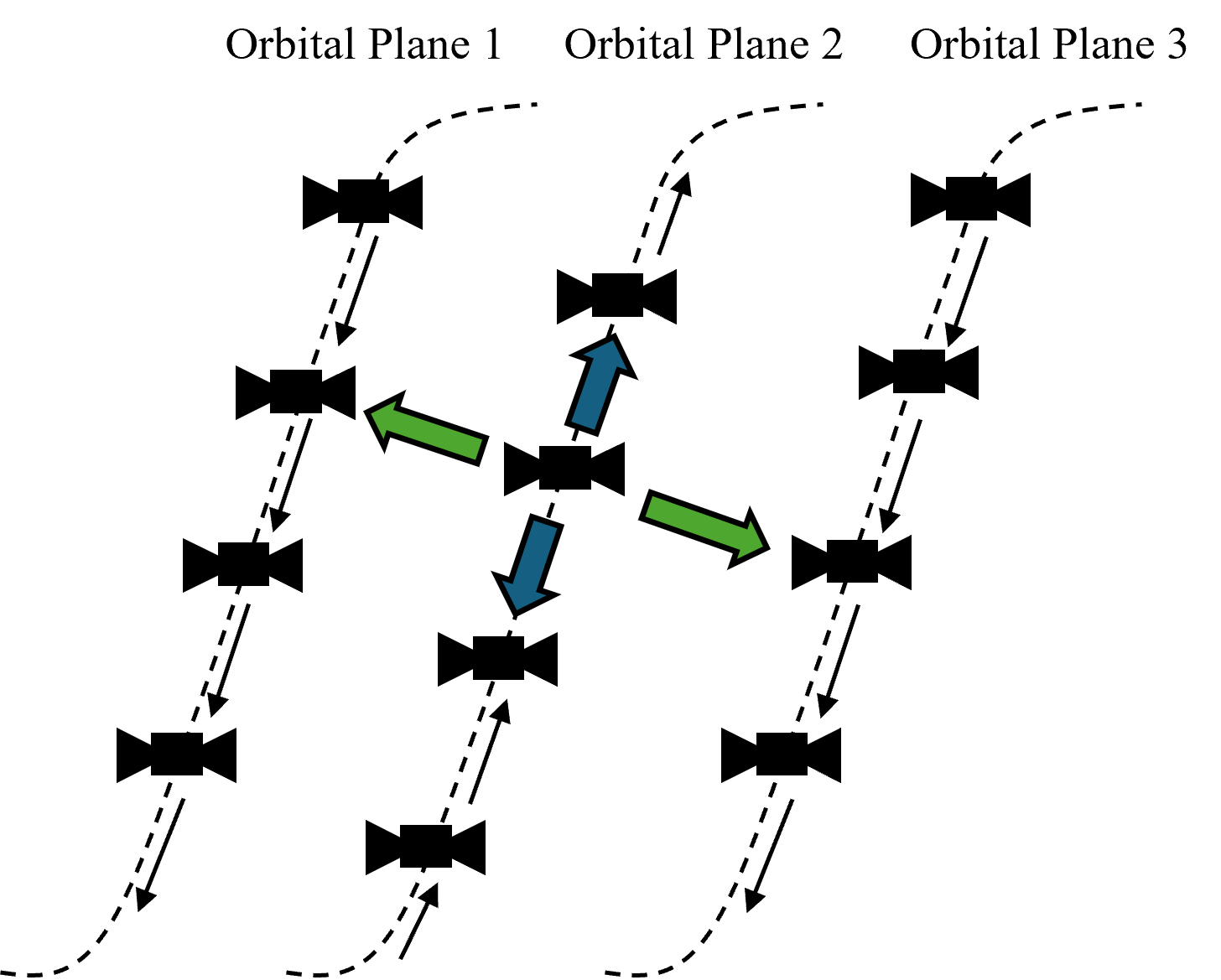}%
        \label{fig:system_b}
    }
    \caption{NGSO system model: (a) ground segment and (b) spacecraft neighborhood.}
    \label{fig:system}
\end{figure*}

We consider a NGSO broadband network built on a Walker constellation with $N_s$ satellites evenly distributed over $P$ orbital planes. Each satellite maintains up to four inter-satellite links (ISLs): two intra–plane (east, west) and two inter–plane (forward, backward), yielding a sparse but connected mesh (Fig.~\ref{fig:system}b). The ground segment comprises gateways (GWs) and user terminals (UTs) geographically distributed worldwide (Fig.~\ref{fig:system}a). Each satellite is equipped with a feeder interface that enables a bidirectional link to a GW or UT when within visibility.

\subsection{Network Graph and Traffic Flows}
Let $\mathcal{N} \triangleq \mathcal{S}\cup\mathcal{G}\cup\mathcal{U}$ denote the set of satellites, gateways, and user terminals, and $\mathcal{L}(t)$ the set of links available at time $t$. The network topology is modeled as a time-varying directed graph
\begin{equation}
\mathcal{G}(t) = \big(\mathcal{N},\,\mathcal{L}(t)\big),
\end{equation}
with $\mathcal{L}(t)=\mathcal{L}_{\text{ISL}}(t)\cup\mathcal{L}_{\text{feeder}}(t)$.  

Traffic flows $\mathcal{F}$ are generated between GWs and UTs, and are forwarded exclusively by satellites. Each flow $f \in \mathcal{F}$ is defined by a source–destination pair $(s_f,d_f)$ and an associated demand $\lambda_f$. The feasible action set of satellite $i$ at time $t$ is
\begin{equation}
\mathcal{A}_i(t) \subseteq \{\text{UT},\text{GW},\text{E},\text{W},\text{Fwd},\text{Bwd}\}.
\end{equation}

\subsection{Delay Model}
For each link $(i,j)\in\mathcal{L}(t)$, the instantaneous per–packet delay is modeled as
\begin{equation}
\delta_{ij}(t) =
\underbrace{\frac{d_{ij}(t)}{c_{\text{eff}}}}_{\text{propagation}} +
\underbrace{\frac{\ell}{R_{ij}(t)}}_{\text{transmission}} +
\underbrace{q_{ij}(t)}_{\text{queueing}},
\end{equation}

where $d_{ij}(t)$ is the propagation distance, $c_{\text{eff}}$ the effective signal speed, $\ell$ the packet size, $R_{ij}(t)$ the service rate, and $q_{ij}(t)$ the queuing delay at node $i$.  
The end-to-end delay for flow $f$ along path $\mathcal{P}_f(t)$ is
\begin{equation}
D_f(t) = \sum_{(i,j)\in \mathcal{P}_f(t)} \delta_{ij}(t).
\end{equation}

\subsection{Routing Optimization Problem}
The routing objective is to minimize the aggregate latency of all flows:
\begin{align}
\min_{\{\mathcal{P}_f(t)\}} \quad & \sum_{f \in \mathcal{F}} D_f(t) 
\label{eq:optobj2}
\end{align}
subject to the following constraints:
\begin{align}
& (i,j) \in \mathcal{L}(t), && \forall (i,j)\in \mathcal{P}_f(t), \label{eq:avail2}\\
& \sum_{f \in \mathcal{F}} \lambda_{f}^{ij}(t) \leq R_{ij}(t), && \forall (i,j)\in\mathcal{L}(t), \label{eq:cap2}\\
& (i,j)\in \mathcal{L}_{\text{feeder}}(t) \Rightarrow \text{visibility}(i,j)=1, \label{eq:cov2}\\
& \sum_{j:(i,j)\in \mathcal{L}(t)} x_{ij}^f(t) - \sum_{j:(j,i)\in \mathcal{L}(t)} x_{ji}^f(t) = b_i^f, \label{eq:flow2}
\end{align}
where $\lambda_{f}^{ij}(t)$ is the rate of flow $f$ on link $(i,j)$, $x_{ij}^f(t)\in\{0,1\}$ indicates whether flow $f$ uses $(i,j)$, and $b_i^f$ is the flow balance ($1$ if $i=s_f$, $-1$ if $i=d_f$, and $0$ otherwise).

\subsection{Complexity and Reinforcement Learning Perspective}
The optimization problem \eqref{eq:optobj2}–\eqref{eq:flow2} is inherently time-varying and non-convex due to the stochastic traffic arrivals $\lambda_f(t)$ and dynamic topology $\mathcal{L}(t)$. Exact solutions are intractable for real-time operation, motivating the adoption of reinforcement learning (RL).  

Each satellite acts as an RL agent that observes a local state $s_t^i$ (queue lengths, link availability, estimated delays, and destination labels) and selects an action $a_t^i\in\mathcal{A}_i(t)$. The agent follows a policy $\pi(a|s)$ that maximizes the expected cumulative reward
\begin{equation}
\max_{\pi} \ \mathbb{E}\!\left[\sum_{t=0}^{T} \gamma^t r_t^i \right],
\end{equation}
with reward function
\begin{equation}
r_t^i = -\alpha \cdot q_t^i - \beta \cdot h_t + \gamma \cdot \mathbb{1}_{\text{delivered}},
\end{equation}
where $q_t^i$ penalizes congestion, $h_t$ represents hop count, and $\mathbb{1}_{\text{delivered}}$ rewards successful packet delivery.  

This RL formulation provides a principled basis for the hybrid table-assisted and DQL-based routing strategy developed in the next section.

\section{Hybrid RL- and Table-Assisted Routing Framework}

Building upon the decision process defined in Section~II, we now present two alternative routing strategies: a fully RL-based decentralized policy and the proposed hybrid table-assisted scheme. The distinction lies in how each approach leverages the state, action, and reward structure to balance exploration and exploitation.

\subsection{Approach 1: RL-Based Routing with Offline Exploration and Online Exploitation}

In the baseline method, each satellite operates as an RL agent. Policies are pre-trained through \emph{offline exploration} in simulated environments to cover diverse topologies and traffic patterns, and subsequently deployed for \emph{online exploitation}. At runtime, the policy $\pi_{\text{RL}}$ maps each observed state $s_t^i$ to an action $a_t^i \in \mathcal{A}_i(t)$ according to
\begin{equation}
a_t^i \sim \pi_{\text{RL}}(\cdot|s_t^i).
\end{equation}
Although this strategy provides adaptability, it remains sensitive to distribution shifts between training and operational conditions. As a result, convergence may still be slow, and route stability can degrade when traffic deviates significantly from the training distribution.

\subsection{Approach 2: Table-Assisted Routing with Selective DQL Adaptation}

The proposed method anchors the exploitation baseline on deterministic routing tables, while retaining RL adaptation only when needed. Each satellite is initialized with a table $\mathcal{T}$, where
\begin{equation}
\pi_{\mathcal{T}}(s_f,d_f) = \mathcal{T}(s_f,d_f),
\end{equation}
gives the next-hop under nominal conditions.  

The hybrid decision policy is then defined as
\begin{equation}
\pi_{\text{hyb}}(s) =
\begin{cases}
\pi_{\mathcal{T}}(s), &
\begin{aligned}[t]
  &\text{if the link to }\\& \pi_{\mathcal{T}}(s) \\
  &\text{is available}
\end{aligned} \\[3pt]
\pi_{\text{DQL}}(s) = \arg\max_{a \in \mathcal{A}(s)} Q_\theta(s,a), &
\text{otherwise.}
\end{cases}
\label{eq:hybrid_policy}
\end{equation}

where $Q_\theta$ is the action–value function of the DQL agent.  

In this setting, RL is invoked only as a fallback mechanism, which drastically reduces the frequency of exploration. Moreover, any exploration occurs in a constrained subspace already initialized by a valid deterministic policy, ensuring faster convergence and stable exploitation.

The hybrid policy can be summarized in Algorithm~\ref{alg:hybrid}, which illustrates how table-assisted forwarding is combined with selective DQL adaptation.

\begin{algorithm}[t]
\caption{Hybrid Table-Assisted Routing with DQL Fallback}
\label{alg:hybrid}
\begin{algorithmic}[1]
\STATE Initialize routing table $\mathcal{T}$ (offline shortest-path computation)
\STATE Initialize DQL agent with parameters $\theta$
\FOR{each packet with destination $d_f$ arriving at satellite $i$}
    \STATE $j \gets \mathcal{T}(s_f,d_f)$
    \IF{link $(i,j)$ is available}
        \STATE Forward packet to $j$
    \ELSE
        \STATE $a^* \gets \arg\max_{a \in \mathcal{A}_i(t)} Q_\theta(s_t^i,a)$
        \STATE Forward packet according to $a^*$
        \STATE Update $Q_\theta$ using observed reward $r_t^i$
    \ENDIF
\ENDFOR
\end{algorithmic}
\end{algorithm}

\subsection{Computational Complexity}

The RL-only strategy requires evaluating $Q_\theta(s,a)$ for all $a \in \mathcal{A}(s)$ at every decision step, with cost
\begin{equation}
\mathcal{O}\big(|\mathcal{A}|\cdot d\big),
\end{equation}
where $d$ is the state dimension. Moreover, continuous online exploration incurs additional learning overhead.  

In contrast, the hybrid scheme performs table lookups with cost $\mathcal{O}(1)$ in nominal conditions. Only when fallback is triggered does it incur the $\mathcal{O}(|\mathcal{A}|\cdot d)$ evaluation. Since fallback events are sparse, the average per-packet decision complexity is significantly reduced, making the method scalable to large constellations.

\section{Performance Evaluation}

\begin{figure}[t]
    \centering
    \includegraphics[width=0.9\linewidth]{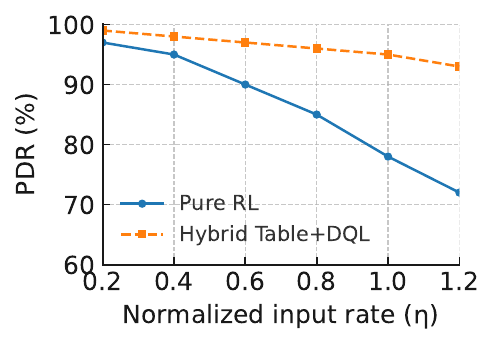}
    \caption{Packet Delivery Ratio (PDR) versus \emph{normalized input rate} $\eta$ (user-offered traffic). The knob $\eta$ scales the per-flow Poisson arrival $\lambda_f=\eta\,\lambda_0$ with $N_F=100$ active flows. The hybrid policy sustains higher PDR across all $\eta$.}
    \label{fig:pdr_load}
\end{figure}

\begin{figure}[t]
    \centering
    \includegraphics[width=0.9\linewidth]{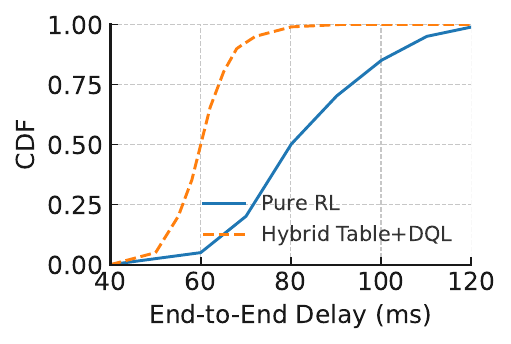}
    \caption{End-to-end delay CDF at $\eta=1$. The hybrid scheme exhibits consistently lower latency and a tighter distribution, evidencing robustness against congestion.}
    \label{fig:delay_cdf}
\end{figure}

\begin{figure}[t]
    \centering
    \includegraphics[width=0.9\linewidth]{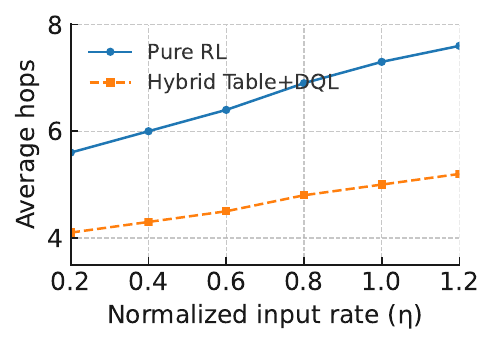}
    \caption{Average hop count versus \emph{normalized input rate} $\eta$. The hybrid policy consistently uses shorter paths across $\eta$, with a widening gap at higher $\eta$.}
    \label{fig:hops_load}
\end{figure}

\begin{figure}[t]
    \centering
    \includegraphics[width=0.9\linewidth]{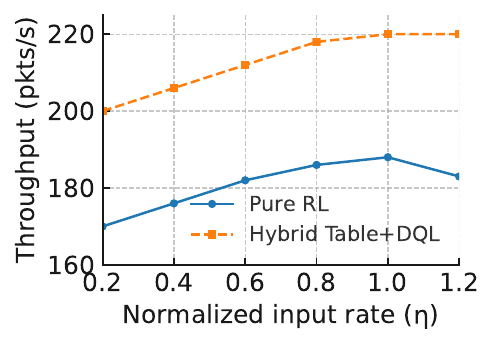}
    \caption{Throughput (carried rate at sinks) versus \emph{normalized input rate} $\eta$. The hybrid approach sustains higher carried rate across all $\eta$ and saturates near $220$ pkts/s, while the pure RL baseline plateaus around $180$--$190$ pkts/s and slightly degrades under saturation.}
    \label{fig:throughput_load}
\end{figure}


\subsection{Simulation Setup}
We use a packet-level MATLAB simulator for an NGSO Walker-Delta constellation. Table~\ref{tab:sim-params} summarizes all parameters. Each simulation runs for $T_{\mathrm{sim}}=600$\,s with a 100\,s warm-up discarded; orbital dynamics are updated every 1\,s. Traffic is GW$\leftrightarrow$UT with Poisson arrivals scaled by the \emph{normalized input rate} $\eta\in[0.2,1.2]$, i.e., $\lambda_f=\eta\,\lambda_0$ for each active flow. Each point averages 20 seeds (seed $=42$). We compare \emph{Pure RL (DQL)} versus \emph{Hybrid Table+DQL} and report PDR, end-to-end delay, average hops, and throughput.

\textbf{Fallback metric.} We record the \emph{fallback activation rate} $p_{\mathrm{fb}}(\eta)$ as the fraction of per-packet routing decisions in which the table next hop is invalid (link down or local queue ratio $>0.7B$) and the DQL fallback is invoked.

\begin{table}[t]
\captionsetup{font=footnotesize}
\caption{Simulation parameters (network, traffic, and learning).}
\label{tab:sim-params}
\centering
\scriptsize
\setlength{\tabcolsep}{3.5pt}
\renewcommand{\arraystretch}{1.0}
\begin{tabular}{p{0.36\linewidth} p{0.58\linewidth}}
\hline
\textbf{Constellation} & Walker-Delta; $P{=}8$, $S{=}20$/plane ($N_s{=}160$); $h{=}550$\,km, $i{=}53^\circ$; uniform RAAN/MA. \\
\textbf{ISLs} & Up to $4$/sat: intra-plane (E/W), inter-plane (Fwd/Bwd); active if distance $<\!2000$\,km; $R_{\text{ISL}}{=}1$\,Gbps. \\
\textbf{Feeder} & Active if elevation $>\!10^\circ$; $R_{\text{feeder}}{=}2$\,Gbps. \\
\textbf{Propagation} & Geometric distance; $c_{\text{eff}}{=}3{\times}10^8$\,m/s. \\
\textbf{Packets/buffers} & $\ell{=}1200$\,B; FIFO per port $B{=}200$ pkts. \\
\textbf{Traffic} & GW$\leftrightarrow$UT; random pairing over $N_{\mathrm{GW}}{=}12$, $N_{\mathrm{UT}}{=}50$ clusters; Poisson with $\lambda_f{=}\eta\lambda_0$, $\eta\!\in\![0.2,1.2]$. \\
\textbf{Time/avg.} & $T_{\mathrm{sim}}{=}600$\,s (warm-up $100$\,s); orbital step $1$\,s; $20$ runs; seed $=42$. \\
\textbf{Table policy} & Offline Dijkstra (weights: mean per-link delay). \\
\textbf{Fallback trigger} & DQL if next hop down or queue ratio $>\!0.7B$. \\
\textbf{RL obs./actions} & Queues, link-availability flags, per-link delays, destination label; actions $\{\text{UT},\text{GW},\text{E},\text{W},\text{Fwd},\text{Bwd}\}$. \\
\textbf{Reward} & $r_t{=}-\alpha q_t-\beta h_t+\gamma\,\mathbb{1}_{\text{delivered}}$. \\
\textbf{DQL (both)} & MLP 128–64 (ReLU); Adam $10^{-3}$, $\gamma{=}0.99$; target upd. $1000$ steps; replay $10^5$; batch $64$; $\epsilon\!:\!1.0{\to}0.05$ in $2{\times}10^5$; pretrain $10^6$ steps; online exploitation. \\
\textbf{Hybrid runtime} & Table lookup $\mathcal{O}(1)$ nominal; DQL $\mathcal{O}(|\mathcal{A}|\cdot d)$ on fallback; one TD update on fallback. \\
\hline
\end{tabular}
\end{table}

\subsection{Results and Discussion}
Table~\ref{tab:results} summarizes the aggregate performance, while Figs.~\ref{fig:pdr_load}--\ref{fig:throughput_load} detail the behavior as the \emph{normalized input rate} $\eta$ increases.

\begin{table}[t]
\centering
\caption{Performance comparison (consistent with Figs.~\ref{fig:pdr_load}--\ref{fig:throughput_load}).}
\label{tab:results}
\footnotesize
\renewcommand{\arraystretch}{1.1}
\begin{tabular}{p{3cm}p{1cm}p{3.2cm}}
\hline
\textbf{Metric} & \textbf{Pure RL} & \textbf{Hybrid Table+DQL} \\
\hline
PDR (\%) across $\eta$ & $72$--$97$ & $93$--$99$ \\
Throughput (pkts/s) across $\eta$ & $170$--$188$ & $200$--$220$ \\
Average hops across $\eta$ & $5.6$--$7.6$ & $4.1$--$5.2$ \\
Median / 95th delay at $\eta{=}1$ (ms) & $80$ / $110$ & $60$ / $72$ \\
\textbf{Fallback activation $p_{\mathrm{fb}}$ (\%)} & --- & $<\!1$ @ $\eta\!\le\!0.4$;\ $3.2$ @ $0.8$;\ $6.9$ @ $1.0$;\ $11.3$ @ $1.2$ \\
\hline
\end{tabular}
\end{table}

Fig.~\ref{fig:pdr_load} shows the PDR versus the normalized input rate $\eta$. The hybrid method consistently exceeds $93\%$ across all $\eta$, whereas the pure RL baseline degrades below $75\%$ at the highest $\eta$, evidencing the robustness of the hybrid policy under stress.

The latency distribution in Fig.~\ref{fig:delay_cdf} (evaluated at $\eta{=}1$) highlights a tighter delay profile for the hybrid scheme: median $\approx 60$\,ms and 95th percentile $\approx 72$\,ms, compared to $\approx 80$\,ms and $\approx 110$\,ms for pure RL. This confirms that table-assisted forwarding reduces queueing and stabilizes end-to-end delay.

As seen in Fig.~\ref{fig:hops_load}, the average hop count for pure RL increases markedly with $\eta$ (from $\sim 5.6$ to $\sim 7.6$), reflecting congestion-induced detours. The hybrid scheme maintains shorter paths (about one hop fewer on average), indicating more efficient next-hop selection as $\eta$ grows.

Finally, Fig.~\ref{fig:throughput_load} reports throughput (carried rate at sinks) versus $\eta$. The hybrid policy sustains higher carried rate at all operating points and saturates near $220$\,pkts/s, whereas the pure RL baseline plateaus around $180$--$190$\,pkts/s and slightly declines under saturation, showing more effective utilization of link resources.

Overall, fully RL-based routing, while adaptive, exhibits instability and inefficiency under heavy input rates. By combining precomputed routes with selective DQL fallback, the proposed hybrid strategy achieves robustness and efficiency: deterministic forwarding ensures stability in nominal conditions, whereas RL adapts to congestion and link unavailability. This dual-mode design is well suited for large-scale, delay-sensitive NGSO deployments.

Beyond accuracy and throughput, we quantify how often the hybrid policy engages learning. The DQL fallback is rarely needed at low input rates ($p_{\mathrm{fb}}<1\%$ for $\eta\le 0.4$) and remains moderate even near saturation ($p_{\mathrm{fb}}\approx 11\%$ at $\eta=1.2$). Hence, most decisions are $\mathcal{O}(1)$ table lookups, and the more expensive DQL evaluations are confined to congestion or link outages, which explains the stability and lower average complexity observed in practice.

\section{Conclusion and Future Work}
We presented a hybrid routing framework for NGSO constellations that couples deterministic table lookups with a DQL fallback invoked only under link unavailability or congestion. Packet-level evaluations on large constellations show consistent gains over a pure RL baseline—higher PDR, lower end-to-end delay, fewer hops, and larger carried throughput—while preserving adaptability. The benefit stems from table lookups dominating most decisions (thus $\mathcal{O}(1)$ average cost) and restricting DQL evaluations to exceptional events, yielding stable performance under realistic dynamics.

Future work will explore Multi-agent coordination for cooperative decisions; continual/transfer learning to reduce retraining under topology/traffic shifts; on-board implementations under hardware constraints; and integration with handover control and cross-layer optimization (e.g., power control and scheduling).

\section*{Acknowledgments}
We would also like to thank Joel Grotz from SES, who advised us on  the current operational method for routing.

\bibliographystyle{IEEEtran}
\bibliography{ref}

\begin{thebibliography}{10}
\providecommand{\url}[1]{#1}
\csname url@samestyle\endcsname
\providecommand{\newblock}{\relax}
\providecommand{\bibinfo}[2]{#2}
\providecommand{\BIBentrySTDinterwordspacing}{\spaceskip=0pt\relax}
\providecommand{\BIBentryALTinterwordstretchfactor}{4}
\providecommand{\BIBentryALTinterwordspacing}{\spaceskip=\fontdimen2\font plus
\BIBentryALTinterwordstretchfactor\fontdimen3\font minus \fontdimen4\font\relax}
\providecommand{\BIBforeignlanguage}[2]{{%
\expandafter\ifx\csname l@#1\endcsname\relax
\typeout{** WARNING: IEEEtran.bst: No hyphenation pattern has been}%
\typeout{** loaded for the language `#1'. Using the pattern for}%
\typeout{** the default language instead.}%
\else
\language=\csname l@#1\endcsname
\fi
#2}}
\providecommand{\BIBdecl}{\relax}
\BIBdecl

\bibitem{9460990}
J.~Barrueco, J.~Montalban, E.~Iradier, and P.~Angueira, ``Constellation design for future communication systems: A comprehensive survey,'' \emph{IEEE Access}, vol.~9, pp. 89\,778--89\,797, 2021.

\bibitem{11062652}
S.~Pizzi, A.~Tropeano, G.~Araniti, and A.~Molinaro, ``Enhancing orbital edge computing through isl-aided federation of satellite constellations,'' \emph{IEEE Open Journal of the Communications Society}, vol.~6, pp. 5829--5839, 2025.

\bibitem{10436098}
Y.~Lyu, H.~Hu, R.~Fan, Z.~Liu, J.~An, and S.~Mao, ``Dynamic routing for integrated satellite-terrestrial networks: A constrained multi-agent reinforcement learning approach,'' \emph{IEEE Journal on Selected Areas in Communications}, vol.~42, no.~5, pp. 1204--1218, 2024.

\bibitem{HUANG2023284}
\BIBentryALTinterwordspacing
Y.~HUANG, S.~WU, Z.~KANG, Z.~MU, H.~HUANG, X.~WU, A.~J. TANG, and X.~CHENG, ``Reinforcement learning based dynamic distributed routing scheme for mega leo satellite networks,'' \emph{Chinese Journal of Aeronautics}, vol.~36, no.~2, pp. 284--291, 2023. [Online]. Available: \url{https://www.sciencedirect.com/science/article/pii/S1000936122001297}
\BIBentrySTDinterwordspacing

\bibitem{10.1108/978-1-83797-052-020241016}
\BIBentryALTinterwordspacing
S.~Goutam and A.~Goutam, ``Recent trends in deployment of multi-protocol label switching (mpls) networks in universities,'' in \emph{Global Higher Education Practices in Times of Crisis: Questions for Sustainability and Digitalization}.\hskip 1em plus 0.5em minus 0.4em\relax Emerald Publishing Limited, 11 2024. [Online]. Available: \url{https://doi.org/10.1108/978-1-83797-052-020241016}
\BIBentrySTDinterwordspacing

\bibitem{10624807}
B.~Soret, I.~Leyva-Mayorga, F.~Lozano-Cuadra, and M.~D. Thorsager, ``Q-learning for distributed routing in leo satellite constellations,'' in \emph{2024 IEEE International Conference on Machine Learning for Communication and Networking (ICMLCN)}, 2024, pp. 208--213.

\bibitem{10624767}
F.~Lozano-Cuadra and B.~Soret, ``Multi-agent deep reinforcement learning for distributed satellite routing,'' in \emph{2024 IEEE International Conference on Machine Learning for Communication and Networking (ICMLCN)}, 2024, pp. 1--2.

\bibitem{10969776}
F.~Lozano-Cuadra, B.~Soret, I.~Leyva-Mayorga, and P.~Popovski, ``Continual deep reinforcement learning for decentralized satellite routing,'' \emph{IEEE Transactions on Communications}, pp. 1--1, 2025.

\bibitem{10551688}
H.~Liu, Y.~Wang, P.~Li, and J.~Cheng, ``A multi-agent deep reinforcement learning-based handover scheme for mega-constellation under dynamic propagation conditions,'' \emph{IEEE Transactions on Wireless Communications}, vol.~23, no.~10, pp. 13\,579--13\,596, 2024.

\bibitem{11000293}
Y.~Zhang, W.~Jiang, P.~Wan, A.~Liu, H.~Han, J.~Mu, S.~Liu, W.~Gu, S.~Huang, and Z.~Feng, ``Building an efficient satellite network routing scheme: Challenges and opportunities,'' \emph{IEEE Communications Standards Magazine}, vol.~9, no.~2, pp. 30--38, 2025.

\end{thebibliography}

\end{document}